\documentstyle[aps,prb,epsfig]{revtex} 
\newcommand{\beq}{\begin{eqnarray}} 
\newcommand{\eeq}{\end{eqnarray}} 
\begin{document} 
\draft \twocolumn 
 
\title 
{Turning Insulators into Superconductors} 
\author{Philip Phillips} 
 
%
\address 
{Loomis Laboratory of Physics,University of Illinois at 
Urbana-Champaign, 1100 W.Green St., Urbana, IL, 61801-3080, 
} 
 
%
\address{\mbox{ }}
\address{\parbox{14.5cm}{\rm \mbox{ }\mbox{ }
The recent observation by Batlogg and colleagues of superconductivity in an organic
field-effect transistor is reviewed.}}
\address{\mbox{ }}
\address{\mbox{ }} 
\maketitle

\columnseprule 0pt 
\narrowtext

In the opening scene of Mike Nichols' 1968 film
``The Graduate,'' Mr. McQuire advises Benjamin Braddock emphatically
that ``Plastics'' is the one word that holds
the key to his future.  While Benjamin ultimately ignores this directive
choosing instead Elaine, this advise has been prophetic
as applications of plastics have now become 
commonplace.  A surprising arena into which plastics 
have ventured is 
the electronics industry.  The key finding
in the late 1970's that certain
types of polymeric materials can conduct electricity\cite{polyacetylene} almost as efficiently
as copper wire paved the way for
light-emmiting diodes (LED), electrical sensors, and even
rechargable batteries
made entirely from plastics.  This astounding success
has sparked general interest in the electronic
properties of organic materials.  One of the key
endeavors has been the search for organic superconductors\cite{little}.
In a superconductor, electrons bind together to form
Cooper pairs.  At a sufficiently low temperature, the pairs
form a coherent state which conducts electricity without resistive
loss.  On page 702 of
this issue\cite{b1}, a team of scientists from Bell Laboratories, Lucent
Technologies led by Bertram Batlogg 
has taken a rather novel approach to the problem of engineering
superconductivity in an organic material:  they have made
a transistor 
from anthracene, tetracene, and pentacene.  Under normal conditions, these simple organic
materials are insulators, but they can be converted into superconductors by
the simple method of injecting them with charge and lowering the temperature.

The materials in this series are composed of stacks 
of chains containing three (anthracene) to five (pentacene) benzene rings.
Between $T=2-4$K, the resistivity was observed to drop to zero.
In addition expulsion of a magnetic field below a well-defined upper
critical field was also observed. Zero resistance and expulsion
of magnetic fields are intrinsic facts about a true superconducting state.

Given the high superconducting transition temperatures ($T_c=135K$) that
have been
 observed in the copper-oxide materials,
 the question arises:  Why should we care about a relatively low $T_c$ in
 an organic transistor?  The answer is quite simple--these materials are
 far from ordinary
 and are likely to constitute a fertile playground for the study
 of the fundamental properties of correlated electrons.
 Three key features set apart the experimental observations reported here:
 1) dimensionality,
 2) doping process, and 3) no superconducting compounds of anthracene,
 tetracene, and pentacene are known to exist.  
 Regarding
 the first, superconductivity
 is generally associated with a bulk material.  However, in these
 experiments,
 electrons are confined to move at the surface of an organic.  As Fig. 1
 of the Bell labs paper illustrates, two-dimensional confinement is
 accomplished by first coating a pure single polyacene crystal
 with an insulator, typically aluminum oxide and second by creating
a field-effect-transistor geomery with source and drain electrodes 
 on the organic crystal and a gate electrode atop the gate oxide.
    
 Doping carriers into the interface is mediated by simply
changing the voltage at the electrical gate.
  A positive (negative) bias voltage on the gate
 attracts electrons (holes) from the plastic to
 the interface with the insulator. Doping via gating
 has the advantage over traditional chemical doping methods
 in that the physical attributes (such as the lattice
 constant) of the nascent material
 remain unchanged. As electrons cannot
 penetrate the insulator, the charges are confined to move solely
 at the planar interfacial region.   The resistivity is measured by source
 and drain electrodes attached
 to opposite sides of the crystal.  Such a device is a
 field-effect transistor (FET) and is identical electrically
 to those made from semiconductors, such as Si and GaAs. In fact,
 at low temperatures (1-2 K), the mobility\cite{b2} reaches $10^4cm^2/Vs$
 and hence is on par with most semiconductor devices.  The key surprise
 here is that we no longer need to overcome the engineering hurdles
 that accompany Si and GaAs technology to create a two-dimensional
 electron gas.  This is significant because in the last 20 years, the two
 dimensional
 electron gas has dominated solid state physics, resulting
 in Nobel prizes in 1985 and
 1998.  Exotic phases such as a highly
 correlated electron liquid state with fractionally-charged
 excitations\cite{laughlin},
 Cu-O layers in copper-oxide superconductors, and more recently
 the transition to a new conducting state\cite{krav} in the dilute limit
 are the primary reasons why the physics of correlated 2D electrons
continues dominate solid state physics.  Many of these exotic
 states can now be studied in an organic FET.  In fact, Batlogg's group has
already observed the fractional quantum-Hall effect in pentacene\cite{b2}.

As with all discoveries of superconductivity, the key question
that must be resolved is what is the physical mechanism responsible
for electron pairing.  This question is particularly
germane in the case of the polyacene transistors because no doped form of 
polyacenes has ever been observed to superconduct.  In this respect,
the observation of superconductivity in the polyacenes is quite distinct
from the similar result reported recently in a $C_{60}$ transistor\cite{b4}, at 
a density corresponding to three electrons per buckyball,
because $M_3C_{60}$ is known to superconduct.  Further, it is relatively well
established\cite{gunnar} that $M_3C_{60}$ conforms to the standard paradigm in which
collective crystal vibrations, phonons, provide a dynamical glue that binds
electrons in pairs.  In the context of the polyacenes, 
the electron-phonon\cite{phonon}
coupling constant has also been calculated.  As shown in Fig. 1,
the electron-phonon
coupling constant falls off inversely with the number of $\pi$-electrons
for a series of organic
compounds including polyacenes and $C_{60}$. 
\begin{figure}
\begin{center}
\epsfig{file=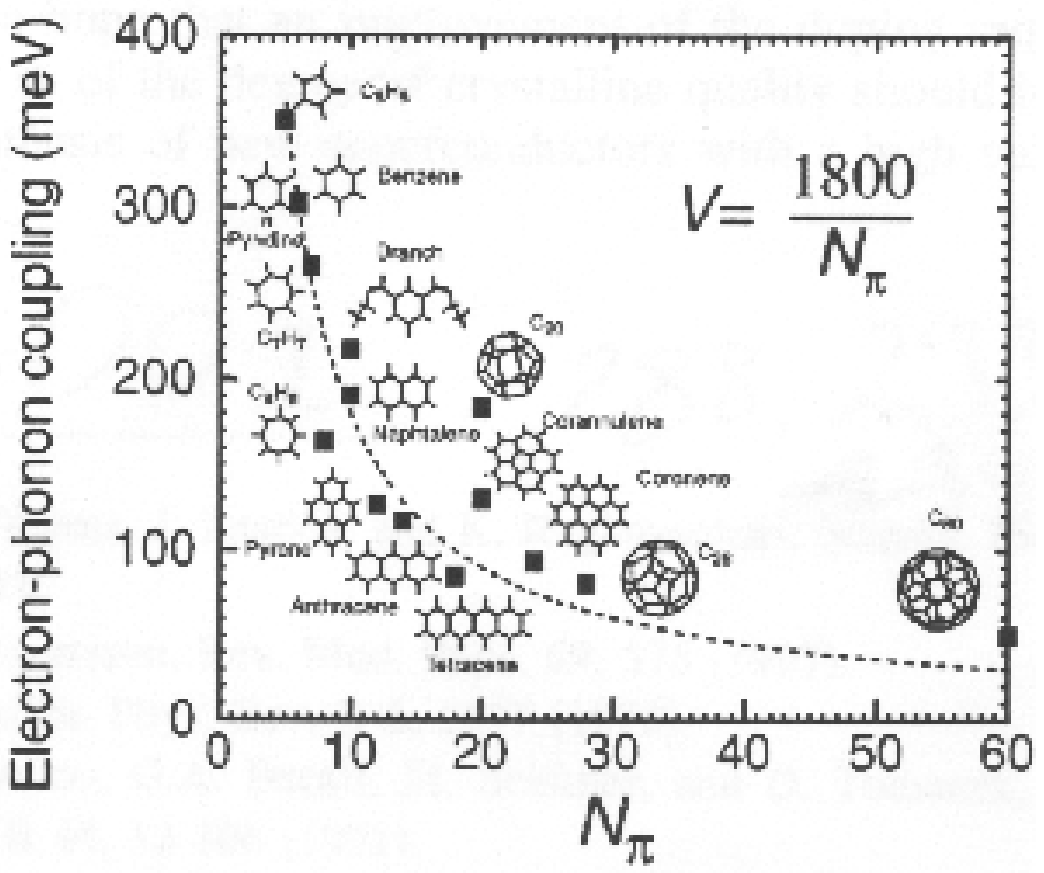, height=6cm}
\caption{What lies behind the superconductivity of organic materials, such
as polyacenes (anthracene) and fullerenes ($C_{60}$)?  A clue might
be provided by the electron-phonon coupling--the `glue' that binds
electrons into superconducting Cooper pairs.  The electron-phonon coupling
constant for a series of organic semiconductors is shown
as a function of the number of $\pi$-electrons associated with their carbon
 rings.
(Redrawn from Ref. 8)  }
\label{fig1}
\end{center}
\end{figure}

Hence, based on this trend, one expects anthracene to have the highest
transition temperature within the polyacene series studied by the Bell
labs group.  Indeed, this is seen experimentally\cite{b1}.  However, also based
on the inverse relationship between the electron-phonon coupling
constant and the number of $\pi$-electrons, one expects doped $C_{60}$ to
have the lowest transition temperature.  In actuality, it has a transition
temperature of $33$K, considerably higher than the 2-4K in the polyacenes.
What might save us here is the density of electronic states and the degeneracy
of the partially-filled molecular level that releases electrons
to the interface.  It is the product of the
density of states and the electron-phonon coupling constant that determines,
to a large extent, the superconducting transition temperature (at least
within traditional mechanisms). However, an
independent measure of the density of states at the Fermi level is 
difficult to achieve experimentally unless thermodynamic measurements are made.
Such measurements might be particularly difficult in the polyacene crystals
because the electron gas is sandwiched between two insulators.  A possible,
though challenging, route
to measuring the density of states is by the construction of a gated
multilayer pentacene structure.  

An additional experiment that would be of utmost importance in determining the
mechanism of pairing would be to probe the properties of a pentacene transistor
at low temperatures and low densities.  Varying the electron density 
changes the strength
of the Coulomb repulsion relative to the kinetic energy of the electrons.
Because the Coulomb repulsion falls off inversely with distance,
whereas the kinetic energy decays as the inverse square of the distance,
the Coulomb repulsion dominates in the low density regime.  Consequently,
it is the low density limit that constitutes the strongly-correlated regime.
While at sufficiently low density
the electrons form a rigid crystalline array, at intermediate densities
there is no consensus as to the precise nature of 
the ground state of the electron gas.
This debate has received renewed fuel recently as a result
of the observation\cite{krav}
of an unexpected conducting state in the dilute regime of a two-dimensional
electron gas in gated Si and GaAs transistors.  There is currently no 
consensus as to the origin
of this state or even if the new conducting state is
genuine.  Because the density can be tuned continuously by simply changing
the gate voltage, the same density regime can be examined in a
pentacene transistor. 

If superconductivity persists in the pentacene
systems at low densities, then this will be truly spectacular but at the same
time problematic.  In the low density regime in which the Coulomb
repulsion greatly exceeds the Fermi energy, the largest kinetic energy
in the system, phonons are too weak to overcome the repulsive interaction
between electrons.  In the dilute regime then, the observation
of superconductivity implies that pairing occurs entirely from repulsive
interactions.  That is, pairing\cite{phillips} is a general organizing principle (or quantum
protectorate to use the language of Laughlin and Pines\cite{lp})
of a dilute two-dimensional electon gas as has been proposed. 
Other more specialized mechanisms involving disorder\cite{bk} and
time-reversal broken states\cite{rz} have also been discussed.  The Bell
labs group is poised to settle this issue definitively.  The problem
raised by a positive answer will be, why hasn't a vanishing resistivity
been observed in Si and GaAs\cite{krav}?  Afterall, if the physics is intrinsic
to the electron gas in the dilute regime, then it should not matter out of
what material the transistor is made.  In light of the  growing 
evidence\cite{mk} that Cooper pairs might possibly exist in
a metallic rather than a true superconducting state
at zero temperature in thin films, resolution of these issues
is not likely to be easy.  Nonetheless, the 
Bell labs innovation offers yet another vehichle for exploring
and hopefully settling some of the unanswered questions
surrounding the correlated states that arise when electrons
are confined to move in a plane.

\end{document}